\documentclass[journal,twoside,web]{ieeecolor}
\let\labelindent\relax
\usepackage[inline]{enumitem}
\usepackage{generic}
\usepackage{cite}
\usepackage{amsmath,amssymb,amsfonts}
\usepackage{algorithmic}
\usepackage{graphicx}
\usepackage{textcomp}
\newtheorem{problem}{Problem}

\def\BibTeX{{\rm B\kern-.05em{\sc i\kern-.025em b}\kern-.08em
    T\kern-.1667em\lower.7ex\hbox{E}\kern-.125emX}}
\begin{document}
\bstctlcite{IEEEexample:BSTcontrol}
\title{Minimally Invasive Live Tissue High-fidelity Thermophysical Modeling using Real-time Thermography}
\author{Hamza El-Kebir, \IEEEmembership{Student Member, IEEE}, Junren Ran, Yongseok Lee, Leonardo P. Chamorro, Martin Ostoja-Starzewski, Richard Berlin, Gabriela M. Aguiluz Cornejo, Enrico Benedetti, Pier C. Giulianotti, and Joseph Bentsman, \IEEEmembership{Senior Member, IEEE}
\thanks{Manuscript submitted for review on July 21, 2022.
Research reported in this publication was supported by the National Institute of Biomedical Imaging and Bioengineering of the National Institutes of Health under award number R01EB029766. The content is solely the responsibility of the authors and does not necessarily represent the official views of the National Institutes of Health. Copyright (c) 2022 IEEE. Personal use of this material is permitted. However, permission to use this material for any other purposes must be obtained from the IEEE by sending an email to pubs-permissions@ieee.org.
}
\thanks{H. El-Kebir is with the Dept. of Aerospace Engineering, University of Illinois Urbana-Champaign, Urbana, IL 61801 USA (e-mail: elkebir2@illinois.edu).}
\thanks{J. Ran, Y. Lee, L. P. Chamorro, M. Ostoja-Starzweski, and J. Bentsman are with the Dept. of Mechanical Science and Engineering, University of Illinois Urbana-Champaign, Urbana, IL 61801 USA (email: \{jran2, yl50, lpchamo, martinos, jbentsma\}@illinois.edu). J.~Bentsman is the corresponding author.}
\thanks{R. Berlin is with the Department of Trauma Surgery, Carle Hospital and Department of Biomedical and Translational sciences, Carle Illinois College of Medicine, Urbana, IL 61801 USA.}
\thanks{G. M. Aguiluz Cornejo, E. Benedetti, and P. C. Giulianotti are with the Dept. of Surgery, University of Illinois at Chicago, Chicago, IL 60612 USA.}
}

\maketitle

\begin{abstract}

We present a novel thermodynamic parameter estimation framework for energy-based surgery on live tissue, with direct applications to tissue characterization during electrosurgery. This framework addresses the problem of estimating tissue-specific thermodynamics in real-time, which would enable accurate prediction of thermal damage impact to the tissue and damage-conscious planning of electrosurgical procedures. Our approach provides basic thermodynamic information such as thermal diffusivity, and also allows for obtaining the thermal relaxation time and a model of the heat source, yielding in real-time a controlled hyperbolic thermodynamics model. The latter accounts for the finite thermal propagation time necessary for modeling of the electrosurgical action, in which the probe motion speed often surpasses the speed of thermal propagation in the tissue operated on. Our approach relies solely on thermographer feedback and a knowledge of the power level and position of the electrosurgical pencil, imposing only very minor adjustments to normal electrosurgery to obtain a high-fidelity model of the tissue-probe interaction. Our method is minimally invasive and can be performed \emph{in situ}. We apply our method first to simulated data based on porcine muscle tissue to verify its accuracy and then to \emph{in vivo} liver tissue, and compare the results with those from the literature. This comparison shows that parameterizing the Maxwell--Cattaneo model through the framework proposed yields a noticeably higher fidelity real-time adaptable representation of the thermodynamic tissue response to the electrosurgical impact than currently available. A discussion on the differences between the live and the dead tissue thermodynamics is also provided.
\end{abstract}

\begin{IEEEkeywords}
Biomedical infrared imaging, thermography, tissue thermodynamics, real-time model estimation.
\end{IEEEkeywords}

\vspace{1cm}

\section{Introduction}
\label{sec:introduction}

\begin{figure*}[t]
    \includegraphics[width=0.68\linewidth]{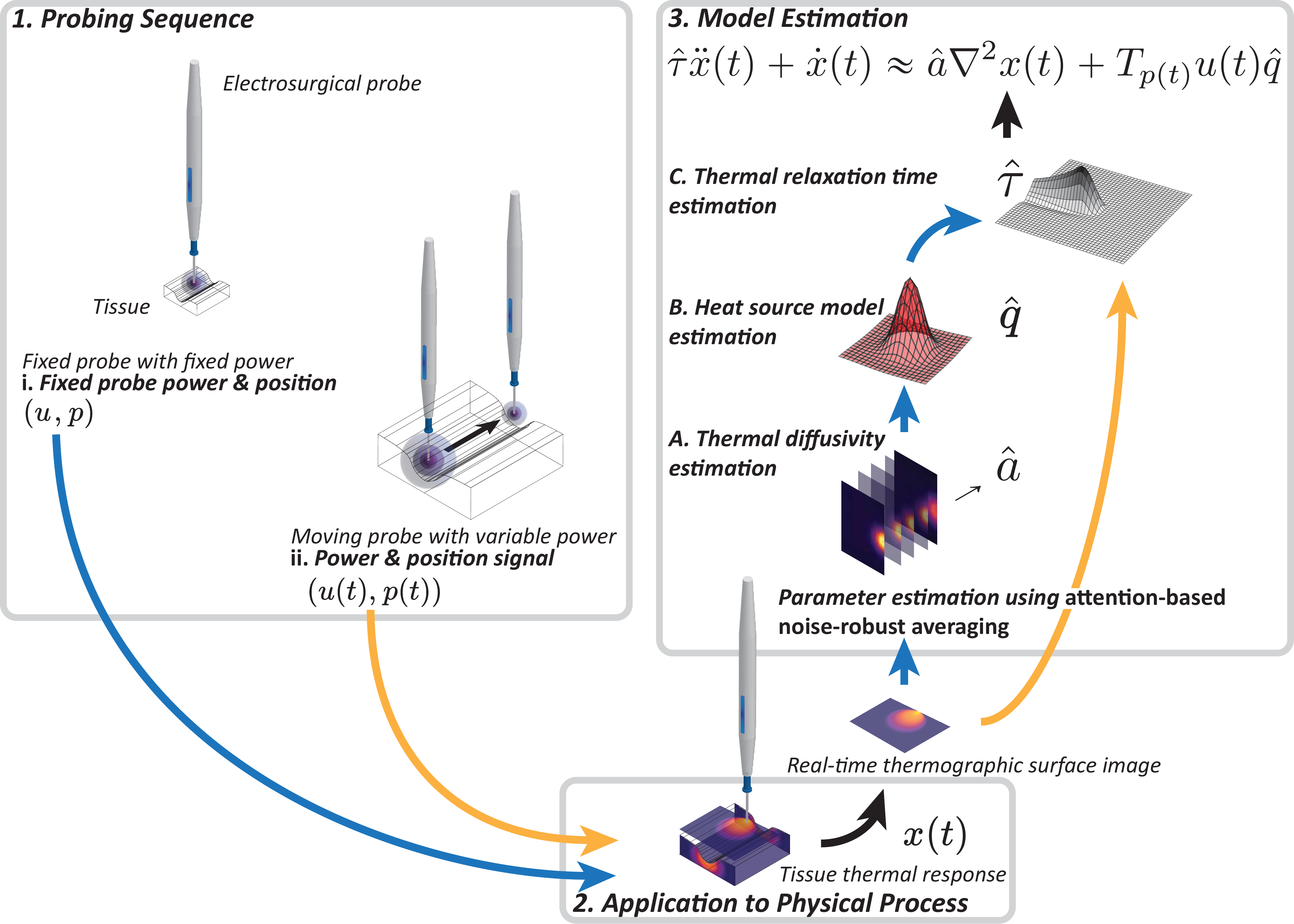}
    \centering
    \caption{Illustration of the proposed parameter estimation for controlled tissue thermodynamics (in this case, electrosurgery is used as an example). In step 1, a probing sequence is determined, consisting of (i) a fixed power stationary probing task, and (ii) a moving cutting task, possibly with variable velocity and power. These probing sequences are applied to the tissue in step 2, during which the tissue thermal response is recorded by a thermographer. These thermographer images are then processed in real-time in step 3, where \emph{attention-based noise-robust averaging} (ANRA) is used to obtain: (A) the thermal diffusivity, (B) the heat source model, and (C) the thermal relaxation time. Subtasks (A) and (B) rely on the stationary probe data, while subtask (C) is based on the moving probe data. This yields an estimated Maxwell--Cattaneo hyperbolic heat transfer model.}
    \label{fig:electrosurgery}
\end{figure*}

\IEEEPARstart{I}{n}
this work, we evaluate the suitability of the hyperbolic Maxwell--Cattaneo model of heat propagation to support a high-fidelity thermodynamic representation of the live tissue response to electrosurgical impact, with applications to porcine liver tissue. As demonstrated in \cite{Madhukar2019, Ran2022}, the movement speeds encountered in electrosurgical procedures (5--20 mm/s), when compared to the speed of heat propagation in tissue (around 2 mm/s), shows that wave-like phenomena due to the finite propagation of thermal information play a dominant role in the overall thermodynamics. An important effect of the slow thermal propagation speed of tissue is the fact that the electrosurgical heat source undergoes distortion from an ideally Gaussian shape to a tear-drop-like shape \cite{Ran2022}, causing significant discrepancies between simulations that use the Maxwell--Cattaneo formulation versus the classical Fourier heat equation in the case of a moving heat source \cite{Ran2021}. Since accurately predicting the thermal spread in live tissues is of prime importance in enabling damage-conscious control of the electrosurgical process and preventing the high heat flux from violating safety margins \cite{El-Kebir2021c, El-Kebir2021d}, we focus in this work on developing a method for estimating the relevant thermophysical parameters for this model based on thermographer readings.

Previous approaches to parameter estimation in service of modeling the electrosurgical process have been undertaken by \cite{Yang2018, Yang2020}, focusing exclusively on estimating the thermal diffusivity, as well as \cite{Madhukar2019}, where the thermal relaxation time was the subject of study. These methods rely on direct solution of the underlying PDEs (partial differential equations), to find a thermal diffusivity that minimizes the error between the simulation result at specific locations that correspond to locations of embedded thermal sensors, and the temperatures recorded at the sensor locations. This method is invasive, requiring embedding sensors into the tissue. In addition, obtaining parameter estimates requires a significant amount of time because the governing PDE must be explicitly solved multiple times to obtain an optimal parameter estimate, precluding the use of such an algorithm for real-time adaptation and safety region mapping during clinical procedures on live tissue. More importantly, multivariate optimization often fails to yield accurate parameter combinations owing to the existence of local minima encountered by optimization algorithms \cite{El-Kebir2022c}. These local minima often necessitate the use of prior knowledge about the parameters to constrain the search space \cite{Yang2018, Yang2020}, thereby inadvertently forcing preconceived biases on the obtained parameter estimates. Our proposed method does not require such assumptions, mainly by a proposed probing sequence that isolates various governing phenomena of the underlying heat transfer problem, allowing parameters to be obtained in series, rather than through the commonly observed error-prone method of simultaneous parameter fitting.

In contrast to the above methods, in \cite{El-Kebir2022c} we presented a method that relies fully on non-invasive remote infrared thermography, in which a thermal camera is used to capture the surface temperature of the tissue. Our approach does not rely on explicitly solving the governing PDEs, but instead employs noise-robust gradient operators to \emph{directly} obtain the PDE's governing parameters by balancing the underlying model differential equations based on the thermographic data. In this work, we build upon these results to achieve a method capable of automatic model estimation for a hyperbolic thermal equation, particularly the Maxwell--Cattaneo heat equation. Central to this endeavor is the adoption of a simple minimally invasive probing sequence, which eliminates ambiguity in the parameter estimation step by isolating the various driving phenomena, namely: diffusion, wave propagation, and external forcing. Our approach allows for \emph{in situ} parameter estimation with a \emph{minimally invasive probing sequence} (obviating biopsies and laboratory testing), with results being obtained in \emph{real-time}. These thermophysical parameters are estimated such that our parsimonious thermodynamics model, the Maxwell--Cattaneo heat equation, best fits the observed thermodynamics. Any unmodeled dynamics are subsumed in this limited set of parameters, producing \emph{effective} parameter estimates that can directly be used for prediction and control of the tissue thermodynamics. To the best of the authors' knowledge, this combination of probing sequence design and direct parameter estimation based on thermographer readings has hitherto not been attempted in the literature. Our method is outlined in Fig.~\ref{fig:electrosurgery}.

This paper makes the following contributions to the existing literature: (i) development of a novel procedure for parameterizing the tissue model in real time under electrosurgical impact, based on thermographic data; (ii) using the latter, enabling a higher-fidelity than currently available modeling of the thermal tissue behavior though parametrizing the Maxwell--Cattaneo tissue model, directly applicable during regular operation without invasive sensor insertion; (iii) based on (ii), offering a high fidelity comparison of tissue thermodynamics under \emph{in vivo} vs \emph{ex vivo} electrosurgical procedures.

This paper is organized as follows. In Sec.~\ref{sec:problem formulation}, the Maxwell--Cattaneo heat transfer model is developed and the parameter estimation problem that will be considered is introduced. Sec.~\ref{sec:parameter estimation method} presents the parameter estimation method, providing justification for our design choices and illustrating considerations in composing probing sequences to separate thermodynamic phenomena. In Sec.~\ref{sec:simulation}, simulation results demonstrating the efficacy of our method compared to a known ground truth are used for verification. In Sec.~\ref{sec:applications}, we apply our method to real-life thermographic data from electrosurgery on \emph{in vivo} porcine liver tissue, discussing differences between the two, and showing improvements in the simulation results based on the updated parameters. Conclusions and future research directions are presented in Sec.~\ref{sec:conclusion}.

\section{Problem Formulation}\label{sec:problem formulation}

As mentioned in the introduction, in this work we seek to obtain a heat transfer model that captures the salient dynamics observed in energy-based surgery methods applied to live tissue. In our model, we wish to include the following phenomena:

\begin{enumerate}[label=\roman*)]
    \item Diffusion (heat conduction);
    \item Presence of a controlled external heat source;
    \item Finite heat propagation time;
\end{enumerate}

The first of these phenomena, diffusion, is by far the most studied in the field of heat transfer in live tissue \cite{Duck1990, Yang2018, Yang2020, Singh2020} (the former two works studying porcine aorta, and the latter considering general animal tissues). The most commonly used formulation is the classical \emph{Fourier heat equation}, which models thermal conductivity. When operating on a one-dimensional domain, the following expression is obtained based on the thermal diffusivity $a$:
\begin{equation}
    \dot{x}(t, \eta) = a \frac{\partial^2 x(t, \eta)}{\partial \eta^2},
\end{equation}
where $x$ denotes the temperature, $t$ the time, and $\eta$ the spatial location.

After adding the heat source, we obtain the following formulation: 
\begin{equation}
    \dot{x}(t, \eta) = a \frac{\partial^2 x(t, \eta)}{\partial \eta^2} + T_{p(t)} u(t) q(\eta),
\end{equation}
In this expression, the heat source is position at location $p(t)$, where $T_{p(t)}$ is an operator that serves to translate the center of the heat source to $p(t)$. The applied heat source is modeled as a unit heat source field $q$ scaled by the input power $u(t)$.

A major drawback in the classical Fourier heat equation, is the fact that changes in the heat source propagate instantaneously throughout the entire domain \cite{Ran2022, Ran2021}. This assumption is unphysical, as heat waves propagate at a finite speed through any physical medium \cite{Madhukar2019, Ran2022, Ran2021}. An important consequence of an infinite heat propagation speed is a typical assumption that the tissue immediately reacts to external thermal excitation, whereas in practice, when an external heat source is applied there is a finite time before the temperature starts rising \cite{Madhukar2019}. In addition, in the case of electrosurgery, one does not observe a rise of temperature in front of the moving needle, even during regular surgical procedures \cite{Madhukar2019}; this also points at the importance of a finite heat propagation time. Introduction of the latter is often achieved by introducing a second time derivative term, so as to induce wave-like behavior. The formulation thus obtained is commonly referred to as the \emph{Maxwell--Cattaneo equation}, or the \emph{telegraph equation}:
\begin{equation}
    \tau \ddot{x}(t, \eta) + \dot{x}(t, \eta) = a \frac{\partial^2 x(t, \eta)}{\partial \eta^2} + T_{p(t)} u(t) q(\eta).
\end{equation}

Here, the second time derivative is scaled by $\tau$, which is known as the \emph{thermal relaxation time}. A so called \emph{second speed of sound} is often obtained by considering both the thermal diffusivity and thermal relaxation time as follows \cite{Ran2022}:
\begin{equation}\label{eq:second speed of sound}
    c = \sqrt{a/\tau}.
\end{equation}

We wish to obtain parameter estimates on a patient-by-patient basis, irrespective of underlying conditions. In light of this goal, we seek a method that is:
\begin{enumerate}
    \item minimally invasive (i.e., no biopsies or unnecessary tissue resection is required);
    \item applicable \emph{in situ}, without the need for external specialized equipment or laboratory testing;
    \item runnable in real-time, with results immediately available.
\end{enumerate}

This yields the following general problem statement, which can be classified as a \emph{parameter estimation problem}:

\begin{problem}\label{problem:parameter estimation}
    Given $N$ surface thermographer readings $\{\hat{x}^{(i)}\}_{i=1}^N$ taken on subsets of $\hat{\Omega}$ with fixed time interval $\Delta t > 0$, as well as control inputs $(u(t), v(t))$, estimate the following parameters of system \eqref{eq:nominal system}:
    \begin{enumerate}
        \item Thermal diffusivity $a$;
        \item Unit heat source $q$;
        \item Thermal relaxation time $\tau$.
    \end{enumerate}
\end{problem}

We now proceed by briefly introducing a well-posed mathematical formulation of our model on which our results in this work will be based.

\subsection{Mathematical Heat Transfer Model}

In this work, we consider the following thermodynamic model on a two-dimensional compact domain $\Omega \subseteq \mathbb{R}^2$:
\begin{equation}\label{eq:nominal system}
    \begin{bmatrix}
        \dot{x}(t) \\
        \ddot{x}(t) \\
        \dot{p}(t)
    \end{bmatrix} =
    \begin{bmatrix}
        0 & 1 & 0 \\
        \frac{a}{\tau} \nabla^2 & - \frac{1}{\tau} & 0 \\
        0 & 0 & 0
    \end{bmatrix}
    \begin{bmatrix}
        x(t) \\
        \dot{x}(t) \\
        p(t)
    \end{bmatrix}
    +
    \begin{bmatrix}
        0 \\
        \frac{1}{\tau} T_{p(t)} u(t) q \\
        v(t)
    \end{bmatrix},
\end{equation}
where $a > 0$ denotes the thermal diffusivity, $\tau > 0$ denotes the thermal relaxation time, and $\nabla^2$ is the Laplacian operator. $x(t)$ denotes the temperature field at time $t$, such that $x(t) \in H^1 (\Omega)$, where $H^1(\Omega)$ refers to the Sobolev space $W^{1,2}(\Omega)$; by $x(t, \eta)$ we denote $x(t)(\eta)$ for $\eta \in \Omega$. The heat source position is denoted by $p(t) \in \Omega$, which is used as a parameter in the translation operator $T$, defined as
\begin{equation}
\begin{split}
    T_{p} : H^1 (\Omega) &\to H^1 (\Omega), \quad \forall p \in \Omega, \\
    T_{p} q(\eta) &\mapsto \chi_{\Omega}(\eta) q(\eta - p), \quad \forall \eta \in \Omega, q \in H^1 (\Omega),
\end{split}
\end{equation}
where $\chi_\Omega$ is the indicator function on the set $\Omega$. In our model, $q \in H^1(\Omega)$ denotes a normalized heat source term, which is scaled by control input $u(t) \in \mathbb{R}_+$, which indicates the electrical power provided by the electrosurgical power supply. Finally, $v \in C^0 (\mathbb{R}^n)$ is the known controlled heat source velocity, which is such that $p(t) := p(0) + \int_0^t v(\tau) \ \mathrm{d}\tau \in \Omega$ for all $t \in [0, \infty)$. We denote the full state by $z \in H^1 (\Omega) \times L^2 (\Omega) \times C^1(\Omega)$, which includes the temperature, the temperature rate of change, and the probe position.

In this work, we assume having access to thermographer data comprised of temperatures that are sampled on a finite Cartesian grid with uniform spacing $\Delta \eta > 0$, denoted by $\hat{\Omega} \subset \Omega$. We consider $\mathrm{dim} \ \hat{\Omega} = M < \infty$, such that $\hat{\Omega} := \{\eta^{(i)}\}_{i=1}^M$.

In the remainder of this work, we make the following assumptions:

\begin{enumerate}[label=(A\arabic*)]
    \item The initial condition $x(0) \in H^1 (\Omega)$ has known constant temperature, i.e., $x(0, \eta) = x_0 \chi_{\Omega}(\eta)$ for all $\eta \in \Omega$ for some known $x_0 \in \mathbb{R}$.
    \item $\inf_t u(t) \geq 0$ and $\sup_t u(t) < \infty$.
    \item On the boundary of $\Omega$, $\partial \Omega$, a homogeneous Neumann boundary condition holds, i.e., $\nabla x(t) |_{\partial\Omega} \equiv 0$.
    \item Heat source velocity $v \in C_\infty^0 (\mathbb{R}^n)$ is continuous and bounded, such that $p(t) := p(0) + \int_0^t v(\tau) \ \mathrm{d}\tau \in \Omega$ for all $t \in [0, \infty)$.
\end{enumerate}

As can be seen in \eqref{eq:nominal system}, we assume that there are no errors in the probe position, nor the heat source model, i.e., \eqref{eq:nominal system} is considered to be a nominal model of the system's thermodynamics, which can be used in a subsequent controller design or for predicting/assessing tissue damage. In the remainder of this work, we consider control input pair $(u(t), v(t))$ to be given for all $t \in [0, \infty)$; we also assume that there is no error in the control inputs.

\paragraph*{Existence and Uniqueness}

To prove the existence and uniqueness of \eqref{eq:nominal system}, we may apply the \textit{Picard iteration method} on the system to obtain a unique solution ($x(t, \eta)$) for known input pair $(u(t), v(t))$ and with any initial condition ($x_0(\eta)$) that satisfies assumption (A1) and (A2) \cite[\S 2.3]{Evans2010}; continuity of the ordinary differential equation (ODE) that defines $p(t)$ follows directly from (A4). Thus, well-posedness of the heat equation under consideration is guaranteed.

\subsection{Probing Sequence for Parameter Estimation}\label{sec:probing seq}

Rather than attempting to estimate all three parameters of system \eqref{eq:nominal system} at once, it bears mentioning that, in general, there is no unique solution to Problem~\ref{problem:parameter estimation} for an arbitrary electrosurgical cutting sequence. Indeed, the ratios $a/\tau$ and $q/\tau$ do not necessarily admit unique solutions without imposing further constraints, especially when considering standard methods that are prone to finding only locally optimal parameters, not to mention the case of unmodeled dynamics and noise when dealing with real-life data. To address this problem, a central idea is to precede the main cutting task by a series of minimally invasive probing tasks, to limit the number of phenomena simultaneously being excited to obtain a unique set of parameters.

Let us rewrite the dynamics of $x(t)$ from \eqref{eq:nominal system}:
\begin{equation}\label{eq:forced moving response}
    \tau \ddot{x}(t) + \dot{x}(t) = a \nabla^2 x(t) + T_{p(t)} u(t) q.
\end{equation}

From observations in \cite{Ran2022}, it follows that a non-moving source with steady heat input results in $\tau \ddot{x}(t)$ vanishing after a transient, whose duration is dictated by thermal relaxation time $\tau$. Therefore, it is possible to study the following equation in isolation, i.e., the tissue's \emph{free response}:
\begin{equation}\label{eq:free response}
    \dot{x}(t) = a \nabla^2 x(t).
\end{equation}

To do so, one would need to sufficiently raise the temperature, e.g., by sustained application of a small amount of electrosurgical action without moving the electrosurgical probe's contact point. Starting from a low or steady-state temperature will lead to $\dot{x}(t) \approx 0$, which does not allow for observing $a \nabla^2 x(t)$. Having raised the temperature, after disengaging the heat source, one will need to wait for the thermal wave caused by the sudden lack of external forcing to have subsided as dictated by the second speed of sound.

It suffices to wait for the final heat wave to propagate outside of the field of view of the thermographer, after which the free response of the tissue can be observed. The latter has been the object of study in \cite{El-Kebir2022c}, in which a novel method known as \emph{attention-based noise-robust averaging} (ANRA) has been developed to solve for the tissue's thermal diffusivity $a$ given the free response.

We can then turn our attention to the stationary forced response, in which we study the time during which heat was actively supplied to the tissue. In this case, provided that a constant non-moving heat source is present, after the initial thermal wave has subsided, we observe a \emph{forced stationary response}, governed by the following dynamics:
\begin{equation}\label{eq:forced response}
    \dot{x}(t) = a \nabla^2 x(t) + T_{p_0} u_0 q,
\end{equation}
where $p_0$ is the fixed position of the probe, and $u_0$ is the constant power supplied. In what follows, we shall discuss our model of $q$, and the method by which we obtain it. It must be noted that $a$ is assumed to be constant throughout the entire cut. The latter assumption is justified, considering that during electrosurgery, undamaged tissue is constantly contacted by the moving probe. In contrast, damaged tissue is often immediately ablated, or is confined to a small region around the cut \cite{Singh2020}. It also bears mentioning that obtaining $a$ and $q$ from the free and forced response can be achieved through a single stationary probing event, in which the tissue is heated (as dictated by \eqref{eq:forced response}) and subsequently allowed to cool down (as governed by \eqref{eq:free response}).

Finally, we may obtain the thermal relaxation time $\tau$ by studying any form of regular cutting action, as governed by \eqref{eq:forced moving response}. The exact method of estimating $\tau$ will be discussed in the following section.

\section{Parameter Estimation Method}\label{sec:parameter estimation method}

In this section, we present our parameter estimation method to solve Problem~\ref{problem:parameter estimation}. We identify three distinct steps:
\begin{enumerate}[label=\Alph*)]
    \item Determining the thermal diffusivity $a$;
    \item Determining the heat source $q$;
    \item Determining the thermal relaxation time $\tau$.
\end{enumerate}
It will become clear that these steps must be performed in order using both the stationary and moving thermal response, as illustrated in Fig.~\ref{fig:electrosurgery}.

Our method relies solely on surface thermographer readings, and thus serves to obtain a model of the apparent thermodynamics of the tissue, as observed on a 2D surface. Extensions to a 3D domain based on 2D observations have been discussed in related work \cite{El-Kebir2022c}, where an observer for 3D tissue temperature has been developed; such an extension is, however, beyond the scope of the current work.

To enable direct estimation of parameters, we leverage a set of noise-robust gradient operators introduced by P.~Holoborodko \cite{Holoborodko2008, Holoborodko2009}; these combine the capabilities of finite difference operators with those of a low-pass filter, allowing for noise-robust gradient computation \cite{El-Kebir2022c}. In essence, our method reduces the parameter estimation problem from an optimization problem over the solution of \eqref{eq:nominal system} to that of balancing differential equations, as will be shown next. Since it was previously hard to obtain gradient field from thermography data on account of high-frequency noise, such gradient-based balancing techniques have not existed before the concept was introduced by the authors in \cite{El-Kebir2022c}.

In the following, we shall denote noise-robust gradient operators by $\hat{\nabla}$ and $\hat{\partial}_t$. We now proceed with the parameter estimation method for the thermal diffusivity.

\subsection{Thermal Diffusivity Estimation}\label{sec:thermal diffusivity est}

We require that \eqref{eq:free response} holds, with a sufficiently large magnitude of $\dot{x}$ to obtain the thermal diffusivity. As mentioned in Sec.~\ref{sec:probing seq}, this is achieved by raising the temperature of the tissue by application of a small amount of sustained electrosurgical action, after which the free response is studied. Focusing on \eqref{eq:free response}, it is apparent that solving for
\begin{equation}\label{eq:thermal diffusivity balance}
    \hat{\partial}_t \hat{x}(t, \eta) = \hat{a}(\eta) \hat{\nabla}^2 \hat{x}(t, \eta)
\end{equation}
at each $\eta \in \hat{\Omega}$, will yield a field $\hat{a}$ as opposed to a single scalar. Many of these parameter estimates will be erroneous, safe for those estimates obtained in regions with a high rate of temperature change \cite{El-Kebir2022c}. This observation has previously led to the development of \emph{attention-based noise-robust averaging} (ANRA) \cite{El-Kebir2022c}, where an \emph{attention layer} is constructed to allow for an average parameter to be obtained. We refer the reader to \cite{El-Kebir2022c} for a full explanation of this method. 

\subsection{Heat Source Estimation}\label{sec:heat source est}

We assume that the unit heat source is modeled as a Gaussian  heat source of the form \cite{El-Kebir2022c}:
\begin{equation}\label{eq:heat source model}
    q(\eta) = q_0 \exp(-\Vert \eta \Vert^2/\rho^2),
\end{equation}
for $\eta \in \Omega$, where $q_0$ is such that $\int_{\Omega} q(\eta) \ \mathrm{d}\eta = 1$, and $\rho$ controls the spread of heat. Also, the heat source scales linearly with the power applied, denoted by $u$ in \eqref{eq:nominal system}. We also introduce some additional notation: we shall call $\ddot{x}$ the `temperature acceleration,' and $\dddot{x}$ the `temperature jerk.' When applied to real-life data, these repeated derivatives are well defined.

Given a diffusivity estimate $\hat{a}$ as obtained from the free response in Sec.~\ref{sec:thermal diffusivity est}, we can now consider the forced response of the tissue, i.e., the time when the tissue was subject to stationary and constant electrosurgical action. In other words, we stipulate that the forced part of the tissue response be obtained by subjecting the tissue to a constant power, with a constant electrosurgical probe position.

To ensure that the forced stationary response model \eqref{eq:forced response} is valid, it is necessary to wait for the initial thermal waves to subside; the modeling errors accrued when not doing so will be demonstrated in due detail in Sec.~\ref{sec:simulation}. Since this time period is dependent on the thermal relaxation time $\tau$, which has not been obtained yet, we elect to study the second time derivative of the observed temperature field instead. In particular, we consider $\hat{\ddot{x}}_{\text{max}}[k] := \max_{i, j} |\hat{\partial}^2_t \hat{x}_{i,j}[k]|$; the rationale for studying this value lies in the fact that a small magnitude of $\ddot{x}$ results in \eqref{eq:forced response} being an accurate approximation of \eqref{eq:forced moving response}. In this work, we wait until the maximum temperature acceleration $\hat{\ddot{x}}_{\text{max}}[k]$ is below 10\% of the all-time maximum temperature acceleration $\hat{\ddot{x}}_{\text{max, total}} = \sup_{k = 1,\ldots,N} \hat{\ddot{x}}_{\text{max}}[k]$, before assuming that \eqref{eq:forced response} is sufficiently valid; this is illustrated in Sec.~\ref{sec:simulation}.

Based on \eqref{eq:forced response}, we define
    $\hat{S}[k] = \hat{\partial}_t \hat{x}[k] - \hat{a} \hat{\nabla}^2 \hat{x}[k]$.

Given that the heat source is located at position $\eta_0 \in \hat{\Omega}$ and has power $u_0 > 0$, we choose to apply nonlinear least-squares fitting to obtain $(q_0, \rho)$ by considering the residual:
\begin{equation}
    r_{i,j}[k] = S_{i,j}[k] - \hat{Q} \exp\left(-\Vert \eta_{i,j} - \eta_0 \Vert^2 / \hat{\rho}^2\right),
\end{equation}
where $(i, j)$ are chosen in a small radius around $\eta_0$; in this work, a region around $\eta_0$ where $S$ assumes positive values is considered. This yields estimates $\hat{q}_0 = \hat{Q}/u_0$ and $\hat{\rho}$, after taking an unweighted time-average over the period following subsidence of the initial thermal wave, as described previously.

\subsection{Thermal Relaxation Time Estimation}\label{sec:thermal relaxation est}

Having obtained $\hat{a}$, $\hat{q}_0$, and $\hat{\rho}$, we may now obtain an estimate for the thermal relaxation time $\tau$. The latter is particularly difficult to observe, since its effects manifest only in the case of transient behavior, such as initial application of electrosurgical power, or when disabling the electrosurgical probe. In steady-state cutting, as is often observed in surgical practice, wave-like effects are rarely observed, given that the process is in a quasi-steady-state where $\ddot{x}$ is of very small magnitude \cite{Ran2022}. Instead, a diffuse shock front can be observed when cutting speeds are sufficiently high, as is commonly the case \cite{Ran2022}; in practice, the resulting Mach cone is so diffuse that $\ddot{x}$ is not of a significant magnitude to be useful in estimation. This latter shock front will be the subject of study in future work.

The equation we wish to solve for is similar to \eqref{eq:thermal diffusivity balance}, but is instead governed by \eqref{eq:forced moving response}:
\begin{equation}\label{eq:thermal relaxation balance}
\begin{split}
    &-\hat{\partial}_t \hat{x}(t, \eta) + \hat{a} \hat{\nabla}^2 \hat{x}(t, \eta) + T_{p(t)} u(t) \hat{q}_0 \exp\left(-\frac{\Vert \eta - p(t) \Vert^2}{\hat{\rho}^2}\right) \\  &= \hat{\tau}(\eta) \hat{\partial}^2_t \hat{x}(t, \eta),
\end{split}
\end{equation}
where we have used the heat source model obtained in Sec.~\ref{sec:heat source est}.

Our approach is mostly identical to that of Sec.~\ref{sec:thermal diffusivity est} (see \eqref{eq:thermal diffusivity balance}), with the exception that the variables are replaced by those appearing in \eqref{eq:thermal relaxation balance}. To obtain a time average, instead of considering the intensity $I[k]$, we elect to impose an event-driven method. Any time we have a sufficiently large change in $u(t)$ (i.e., $\partial_t |u(t)| > \dot{u}_{\text{threshold}} > 0$), we consider the time starting at that change, up until the time when the third time-derivative of the temperature, $\hat{\dddot{x}}_{\text{max}}[k] := \max_{i, j} |\hat{\partial}^3_t \hat{x}_{i,j}[k]|$, is a fraction of the highest recorded value (in this work, 10\% is considered). This rate of acceleration is also known as \emph{jerk}, a commonly encountered concept in mechanics. The latter condition based on the maximum absolute jerk $\hat{\dddot{x}}_{\text{max}}[k]$ provides a clear means of delineating the time during which the effects of a sudden change are apparent; this is illustrated in Sec.~\ref{sec:simulation}.

\section{Simulation}\label{sec:simulation}

To verify the efficacy of the approach outlined in Sec.~\ref{sec:parameter estimation method}, we simulated \eqref{eq:nominal system} based on the properties of porcine liver tissue. These parameters have been obtained from the literature \cite{Madhukar2019, Ran2022, Duck1990}. They are as follows: a thermal diffusivity $a$ of $0.144$ mm$^2$/s, a thermal relaxation time $\tau$ of 0.1 s, and a heat source spread $\rho$ of 0.5 mm. In all of the simulations, the heat source has a power of 30 watts when engaged.

The problem domain is a 4 by 6 cm grid, with a grid resolution of 0.2 mm; a discretization time step of 2 ms is employed. We consider two data sets: one in which the heat source is stationary (to obtain $a$ and $q$), and the other in which the heat source is moving (to obtain $\tau$). The movement speed is set at 20 mm/s. In all cases, the simulation time is 2 seconds, and the probe is active for 1 second in the stationary case, and 1.5 seconds in the moving case. The simulation was carried out using a custom simulation code based on a finite difference scheme, as described in \cite{Ran2022}.

\subsection{Thermal Diffusivity}

\begin{figure}[t]
    \centering
    \includegraphics[width=0.9\linewidth]{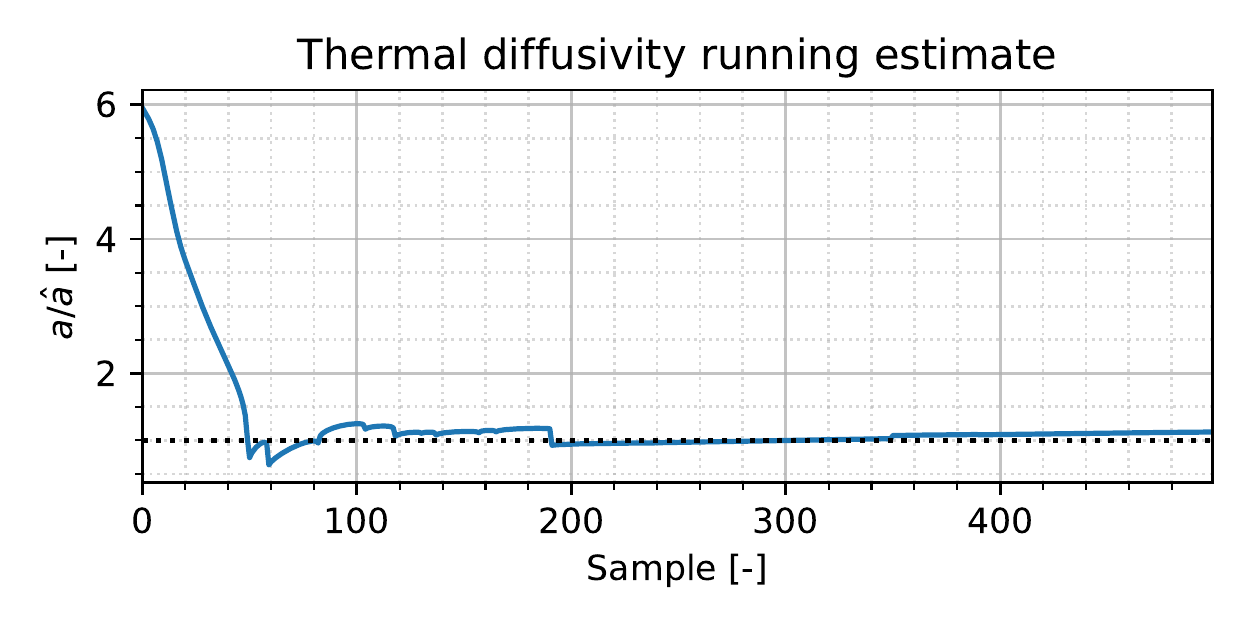}
    \caption{Running estimate of the ratio between the ground truth thermal and the estimated diffusivity, obtained during the unforced part of the stationary thermal response. The dotted horizontal line corresponds to a perfect estimate.}
    \label{fig:sim diffusivity running estimate}
\end{figure}

Employing the method introduced in Sec.~\ref{sec:thermal diffusivity est}, we focus on the last 1 second portion of the stationary simulation run. A cumulative estimate of the diffusivity is shown in Fig.~\ref{fig:sim diffusivity running estimate}, expressed as the ratio $\hat{a}/a$, where the estimate is based on the samples obtained until that instant. The estimated thermal diffusivity $\hat{a}$ is $0.102$ mm$^2$/s, which deviates 11\% from the ground truth. As one may observe, the estimate quickly converges to the ground truth within  100 samples (0.2 s), showing little response to perturbations over time.

\subsection{Heat Source}

As mentioned in Sec.~\ref{sec:heat source est}, the heat source must only be estimated after the initial heat wave subsides, as illustrated by $\hat{\ddot{x}}_{\text{max}}$. We find that the sampling index at which magnitudes below 10\% of the maximum acceleration are maintained is 250 (0.5 s).

\begin{figure}[t]
\begin{minipage}[b]{\linewidth}
  \centering
  \centerline{\includegraphics[width=0.9\linewidth]{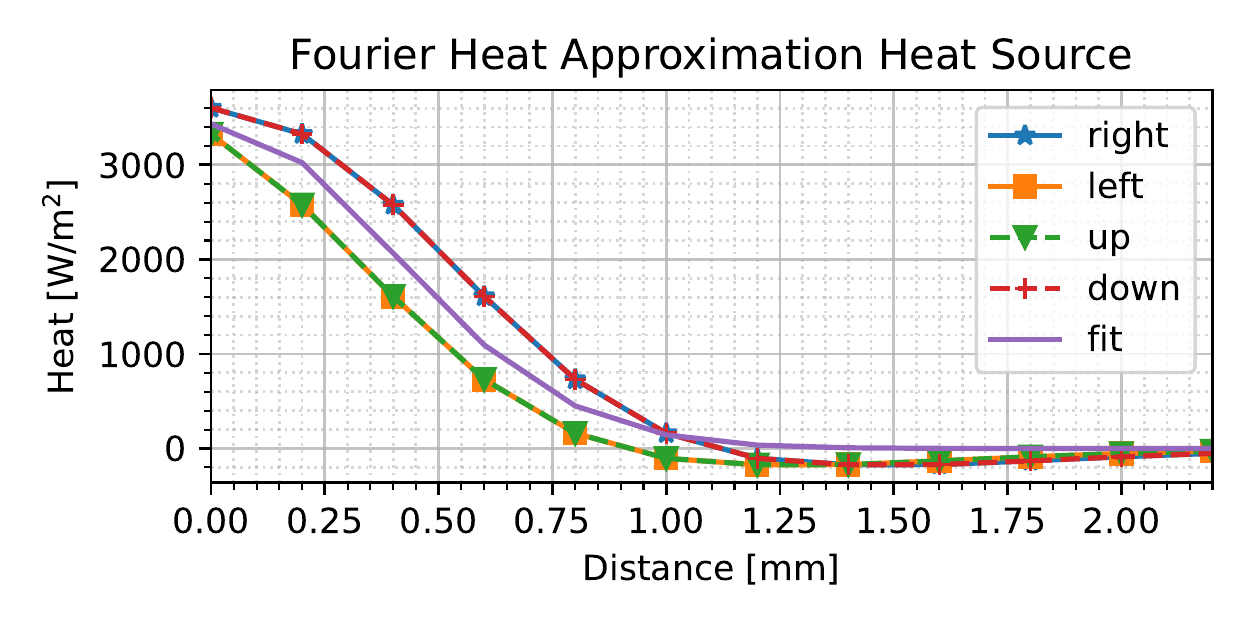}}
  \centerline{\small (a) Estimated Fourier heat source.}
\end{minipage}
\hfill
\begin{minipage}[b]{\linewidth}
  \centering
  \centerline{\includegraphics[width=0.9\linewidth]{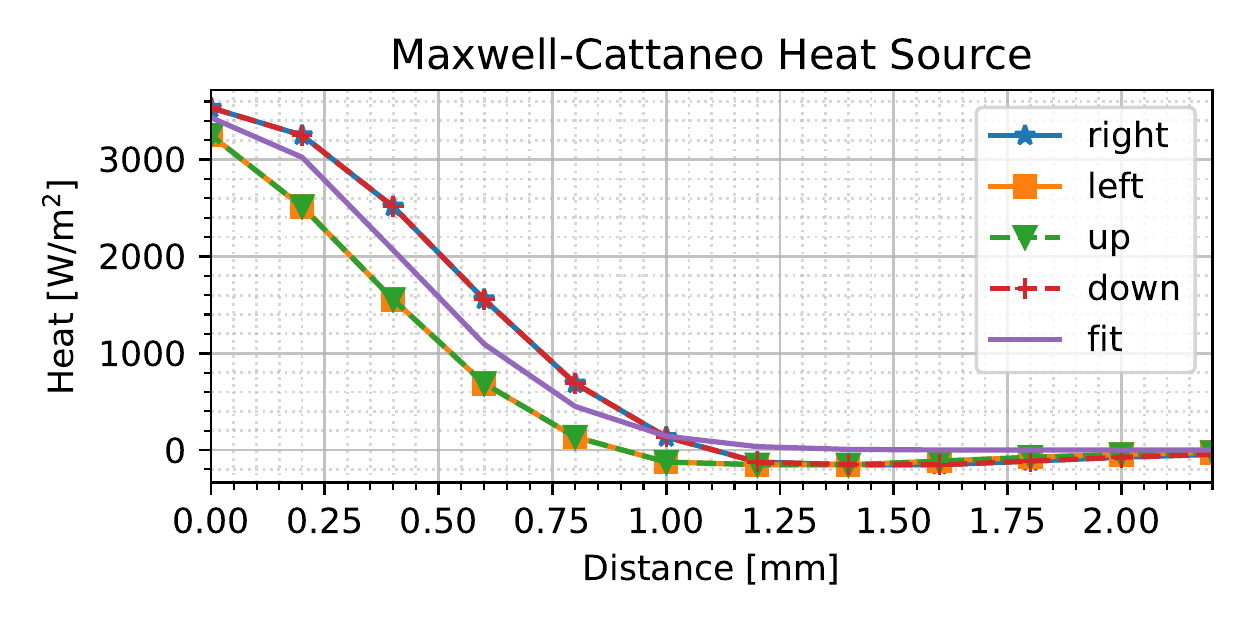}}
  \centerline{\small (b) Ground truth Maxwell--Cattaneo heat source.}
\end{minipage}
    \caption{Heat source model fit compared to (a) Fourier heat source assumption used in fitting the model, (b) ground truth Maxwell--Cattaneo heat source model, \emph{after} the initial heat wave has subsided. The solid line labeled `fit' denotes the model, whereas the other marked lines are the heat fluxes in each of the four cardinal directions. Almost no discrepancies can be observed among the two models.}
    \label{fig:sim heat source correct}
\end{figure}

We apply our heat source fitting method at index 250, after the thermal wave has subsided, yielding the results shown in Fig.~\ref{fig:sim heat source correct}. Here, we have shown the heat source fit with respect to the available data based on \eqref{eq:forced response} (i.e., a Fourier heat equation assumption); we have compared this data to the actual case of the Maxwell--Cattaneo model, in which we have considered \eqref{eq:forced moving response} using the ground truth value of $\tau = 0.1$ seconds. As can be seen in Fig.~\ref{fig:sim heat source correct}(b), we have obtained an adequate fit with respect to the unknown Maxwell--Cattaneo-based heat source. We may assess the accuracy of our model by considering the spread $\rho$ in \eqref{eq:heat source model}, which is obtained to be 0.56 mm, as opposed to the ground truth 0.5 mm.


\begin{figure}[t]
\begin{minipage}[b]{\linewidth}
  \centering
  \centerline{\includegraphics[width=0.9\linewidth]{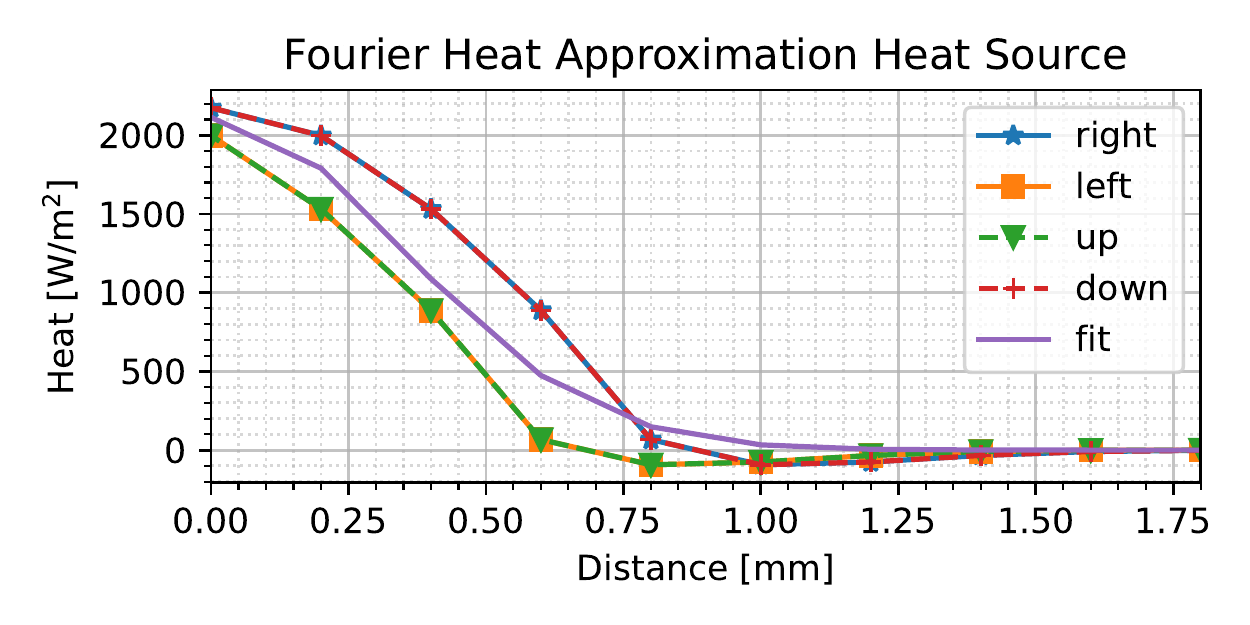}}
  \centerline{\small (a) Estimated Fourier heat source.}
\end{minipage}
\hfill
\begin{minipage}[b]{\linewidth}
  \centering
  \centerline{\includegraphics[width=0.9\linewidth]{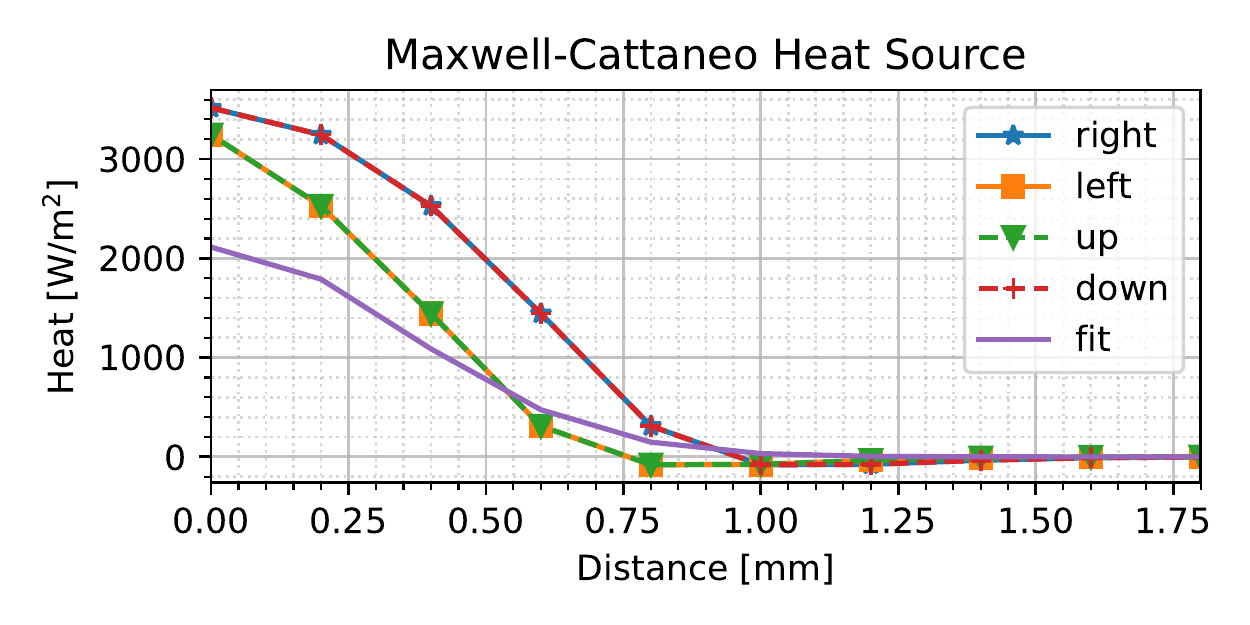}}
  \centerline{\small (b) Ground truth Maxwell--Cattaneo heat source.}
\end{minipage}
    \caption{Heat source model fit compared to (a) Fourier heat source assumption used in fitting the model, (b) ground truth Maxwell--Cattaneo heat source model, \emph{before} the initial heat wave has subsided. The solid line labeled `fit' denotes the model, whereas the other marked lines are the heat fluxes in each of the cardinal directions. Significant discrepancies can be observed owing to the large thermal acceleration.}
    \label{fig:sim heat source incorrect}
\end{figure}

To demonstrate what the results would be if we considered the earlier stages of the forced response when the thermal acceleration is still high (i.e., the thermal wave has not subsided), we have developed the same estimates as in Fig.~\ref{fig:sim heat source correct} at index 50 (0.1 s), as shown in Fig.~\ref{fig:sim heat source incorrect}.

\subsection{Thermal Relaxation Time}

\begin{figure}[t]
    \centering
    \includegraphics[width=0.9\linewidth]{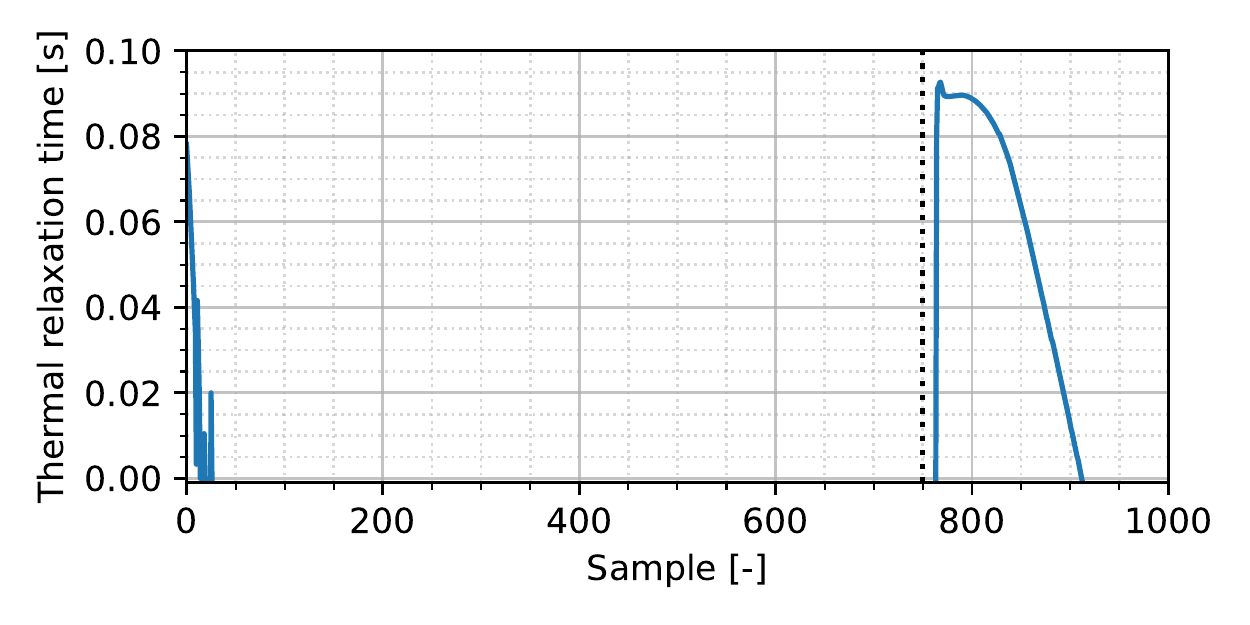}
    \caption{Thermal relaxation time estimate for the moving heat source experiment. The dotted vertical line indicates the time instant at which the heat source is disengaged.  The sampling rate is 28 Hz.}
    \label{fig:sim thermal relaxation estimate}
\end{figure}

To obtain the thermal relaxation time $\tau$, we apply the method introduced in Sec.~\ref{sec:thermal relaxation est} on the moving probe simulation. At each time index, we have given the estimated thermal relaxation constant, as shown in Fig.~\ref{fig:sim thermal relaxation estimate}. Clearly, many of the values are erroneous, except for the initial few values, and more accurately, the values that appear shortly after the heat source is disengaged (as indicated by the dotted vertical line).
Following the averaging criteria given in Sec.~\ref{sec:thermal relaxation est}, we obtain an estimate of $\hat{\tau} = 0.09$ s, deviating 10\% from the ground truth $\tau = 0.1$ s.

%
\vspace{0.5cm}

\section{Application}\label{sec:applications}

In this section, we apply our proposed method to the thermal response of \emph{in vivo} porcine liver tissue to electrosurgical excitation. The trial was performed at the University of Illinois at Chicago's (UIC) Surgical Innovation Training Laboratory (SITL), in accordance with NIH Vertebrate Animal Section standards. The specimen
was provided by the Biological Resources Laboratory of the UIC. After exposing the liver tissue, a stationary point probe was performed using a custom robotic platform based on a Franka Emika Panda robotic arm, equipped with visual and thermal sensors (four Optris Xi 400 thermographers (Optris GmbH, Berlin, Germany)). The electrosurgical generator used was a Valleylab Force FX (Valleylab Inc., CO, USA), applying the monopolar pure cut mode at a power of 5 watts. For the stationary cut, a Covidien Valleylab 2.8" needle electrode (Medtronic Inc., MN, USA) was inserted at 5 mm depth.

\begin{figure}[t]
    \centering
    \includegraphics[width=0.9\linewidth]{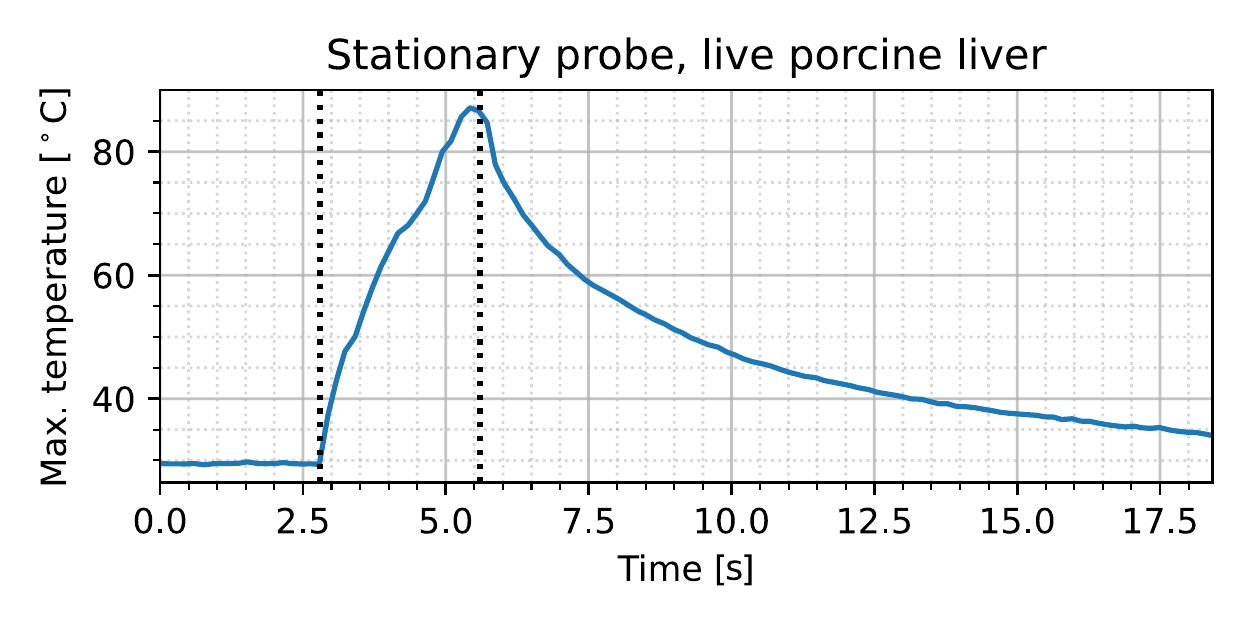}
    \caption{Maximum temperature recorded during the stationary cut trial on live porcine liver tissue. At 2.8 seconds, the power source is engaged, with disengagement taking place at 5.6 seconds, as indicated by the dotted black lines.}
    \label{fig:live max temperature}
\end{figure}

To illustrate the various segments of the stationary probe experiment, the maximum temperature recording during the trial over time is shown in Fig.~\ref{fig:live max temperature}. During the interval 2.8--5.6 seconds, the electrosurgical unit was engaged. Similar to the procedures in Sec.~\ref{sec:simulation}, the period during which the diffusivity was estimated was chosen to be 6.2--12.4 seconds based on the second derivative of temperature. Applying the same method as in Sec.~\ref{sec:thermal diffusivity est}, we obtain at thermal diffusivity of $0.222$ mm$^2$/s, which is about 53\% higher than the non-perfused ex vivo value reported in \cite[Tab.~2.3, p.~16]{Duck1990} \cite{Valvano1984}. This difference likely stems from the presence of blood perfusion, which causes an increase in effective diffusivity of 1.4--1.5 \cite[Sec.~2.1.5.2]{Duck1990} , consistent with our result of 1.53 here.


\begin{figure}[t]
    \centering
    \includegraphics[width=0.9\linewidth]{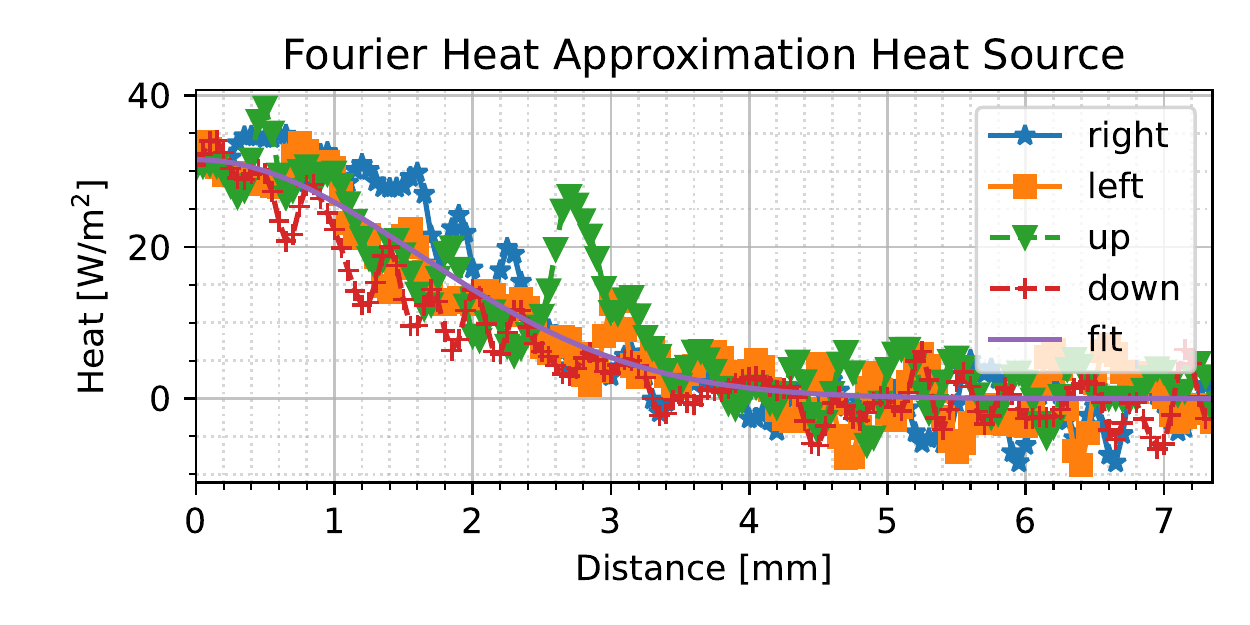}
    \caption{Live porcine liver heat source model fit compared to Fourier heat source assumption used in fitting the model after the initial heat wave has subsided. The solid line labeled `fit' denotes the model, whereas the other marked lines are the heat fluxes in each of the four cardinal directions. Almost no discrepancies can be observed among the two models.}
    \label{fig:live heat source correct}
\end{figure}

Having obtained the thermal diffusivity, we may now estimate the heat source model. We delineate the period during which the heat source model is estimated based on the maximum absolute temperature acceleration criterion (Sec.~\ref{sec:heat source est}), obtaining a time period of interest between 3.8--4.8 seconds. In this case, the heat source estimate field is much more spurious when compared to Sec.~\ref{sec:simulation} as shown in Fig.~\ref{fig:live heat source correct}, but most of the noise has been suppressed by the low-pass properties of the noise-robust gradient operators. In the line corresponding to the `up' direction, however, one may notice an increase in heat flux around 2.75 mm; this anomaly can be attributed to thermal sensor noise, as well as the presence of particulates (smoke), which inevitably introduce discrepancies. Hence, we have based our heat source model on the average heat flux across multiple directions.

To assess the quality of the heat source, we note that our estimated spread is $\hat{\rho} = 2.26$ mm, which is reasonable considering the diameter of the probe (1 mm). The correctness of this result will further be discussed in Sec.~\ref{sec:comparison}.



\begin{figure}[t]
    \centering
    \includegraphics[width=0.9\linewidth]{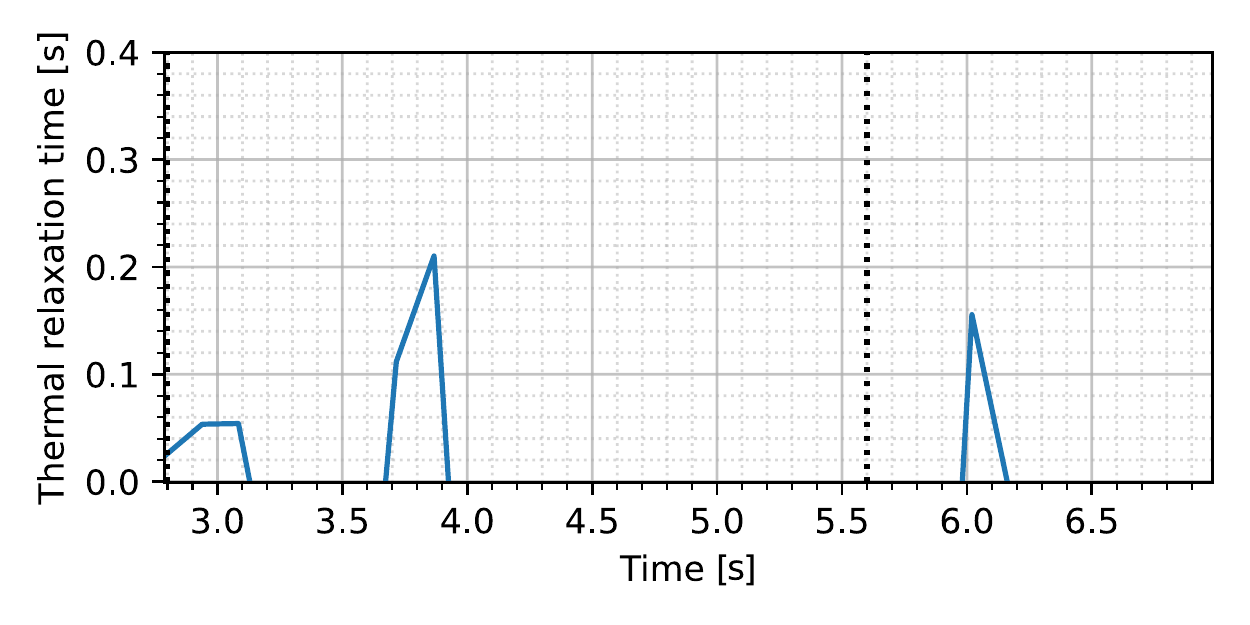}
    \caption{Thermal relaxation time estimate for the stationary heat source experiment on live porcine tissue. The dotted vertical lines indicate the times at which the heat source is engaged and disengaged, respectively.}
    \label{fig:live thermal relaxation estimate}
\end{figure}

We now consider obtaining the thermal relaxation time. Due to issues of image alignment as a result of tissue motion caused by breathing and uncertainties in the thermographer location, we have chosen to estimate the thermal relaxation time based on the stationary probe response, rather than a moving probe response. In an ideal situation, the latter method would yield more conclusive results, since the change in heating may be much more substantial when compared to what can be achieved by stationary probing, while minimizing unnecessary tissue damage. While tissue motion rejection will be the subject of future study, in this work we show that an adequate estimate may also be obtained based on the stationary response. Applying the method of Sec.~\ref{sec:thermal relaxation est}, we find relaxation times as shown in Fig.~\ref{fig:live thermal relaxation estimate}. In the latter figure, we find that the initial engagement of the probe results in two spikes in the thermal relaxation constant estimate. This is a result of the probe temporarily losing contact with the tissue after initial ablation, after which it reconnects with the remaining tissue. This effects can also be observed in the maximum temperature upon close inspection of Fig.~\ref{fig:live max temperature}, where there is a short period of stagnation at the 3.5 second mark. After averaging the relaxation time as described in Sec.~\ref{sec:thermal relaxation est}, we find a relaxation time of $\hat{\tau} = 0.16$ s. This yields a second speed of sound of 1.2 mm/s by \eqref{eq:second speed of sound}, which is close to the results reported in \cite{Madhukar2019, Ran2022} of around 2 mm/s.
The finite heat propagation time can also be observed in the delay between the times where usable estimates of $\tau$ are available in Fig.~\ref{fig:live thermal relaxation estimate} (around $2\tau$ seconds).

\begin{figure}[t]
\begin{minipage}[b]{\linewidth}
  \centering
  \centerline{\includegraphics[width=0.9\linewidth]{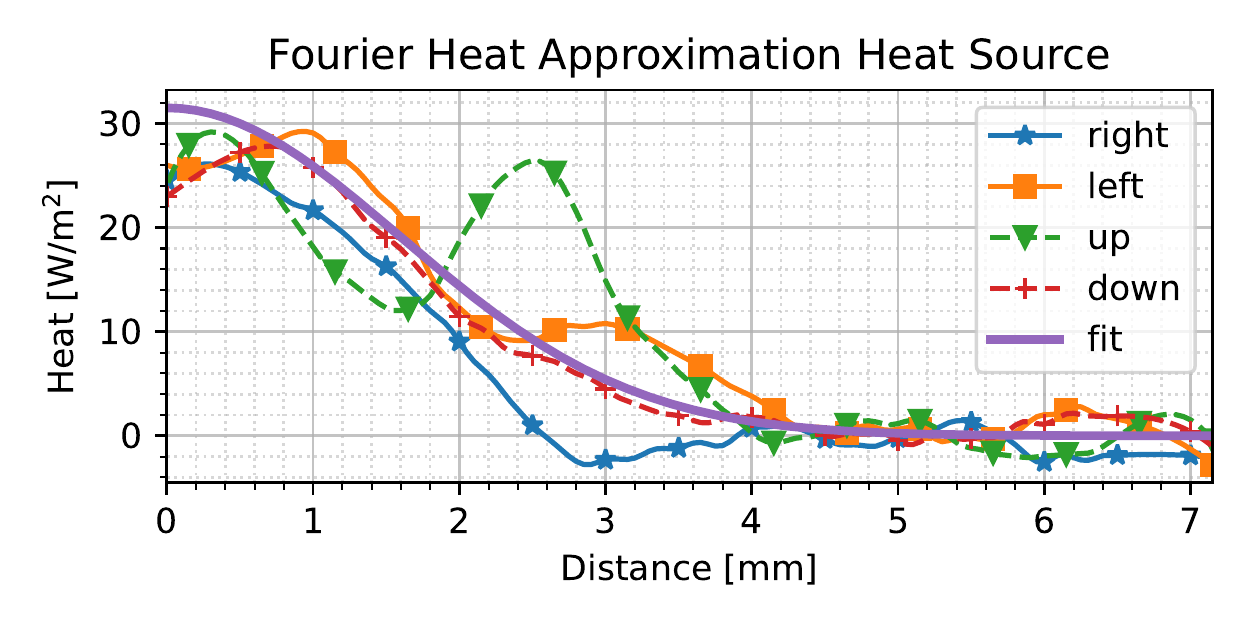}}
  \centerline{\small (a) Estimated Fourier heat source.}
\end{minipage}
\hfill
\begin{minipage}[b]{\linewidth}
  \centering
  \centerline{\includegraphics[width=0.9\linewidth]{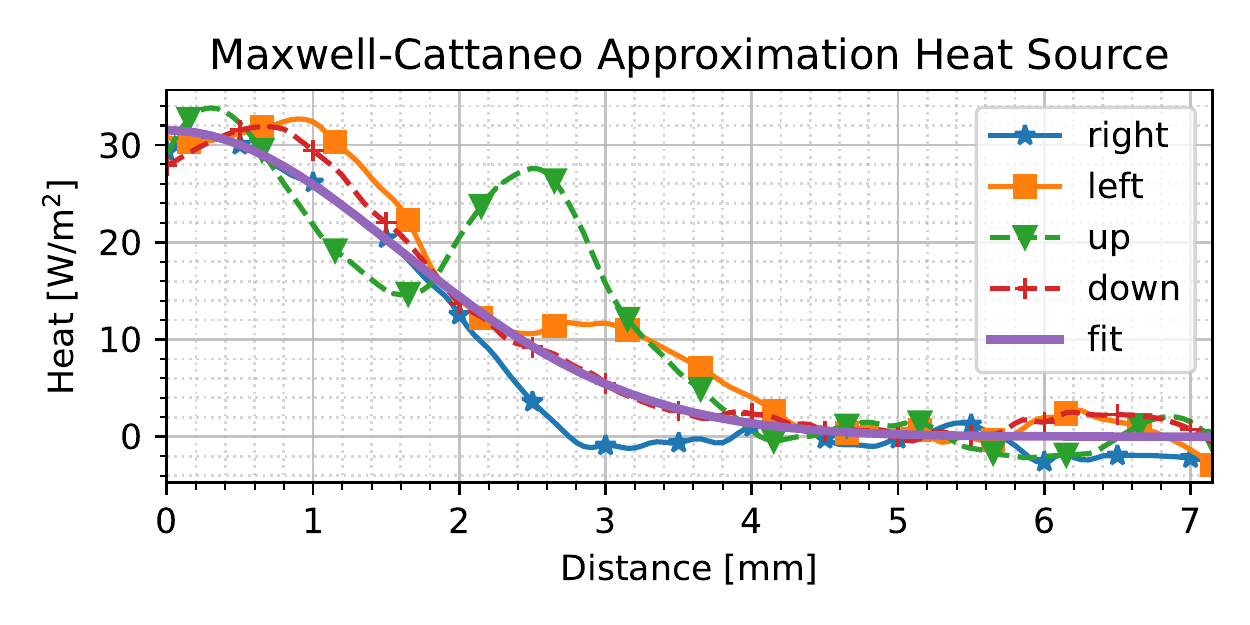}}
  \centerline{\small (b) Expected Maxwell--Cattaneo heat source.}
\end{minipage}
    \caption{Heat source model fit compared to (a) Fourier heat source assumption used in fitting the model, (b) ground truth Maxwell--Cattaneo heat source model, \emph{before} the initial heat wave has subsided for the live porcine liver response. The observed heat flux curves have been smoothened using a Savitzky--Golay filter for ease of viewing. The solid line labeled `fit' denotes the model, whereas the other marked lines are the heat fluxes in each of the cardinal directions. Significant discrepancies can be observed owing to the large thermal acceleration in (a), especially close to the point of application (0 mm distance).}
    \label{fig:live heat source incorrect}
\end{figure}

We may observe the effect of neglecting the thermal relaxation constant based on the error between the Fourier-based and Maxwell--Cattaneo-based heat source approximations with the thermal relaxation time being taken as 0.16 s, as shown in Fig.~\ref{fig:live heat source incorrect}. For ease of viewing, we have smoothened the raw heat flux curves using a third-order Savitzky--Golay filter, with a window size of 31 samples. Clearly, \emph{the Fourier heat approximation exhibits a large offset close to the probe location}, where an error of 10 W/m$^2$ is observed in Fig.~\ref{fig:live heat source incorrect}(b) at the point of application (0 mm distance); \emph{such an error will yield poorly agreeing simulation results across the entire domain}. In the next subsection, we will show that our heat model produces an adequate response when compared to the real-life thermal measurements.

\subsection{Comparison with Simulation}\label{sec:comparison}

\begin{figure}[t]
    \centering
    \includegraphics[width=0.9\linewidth]{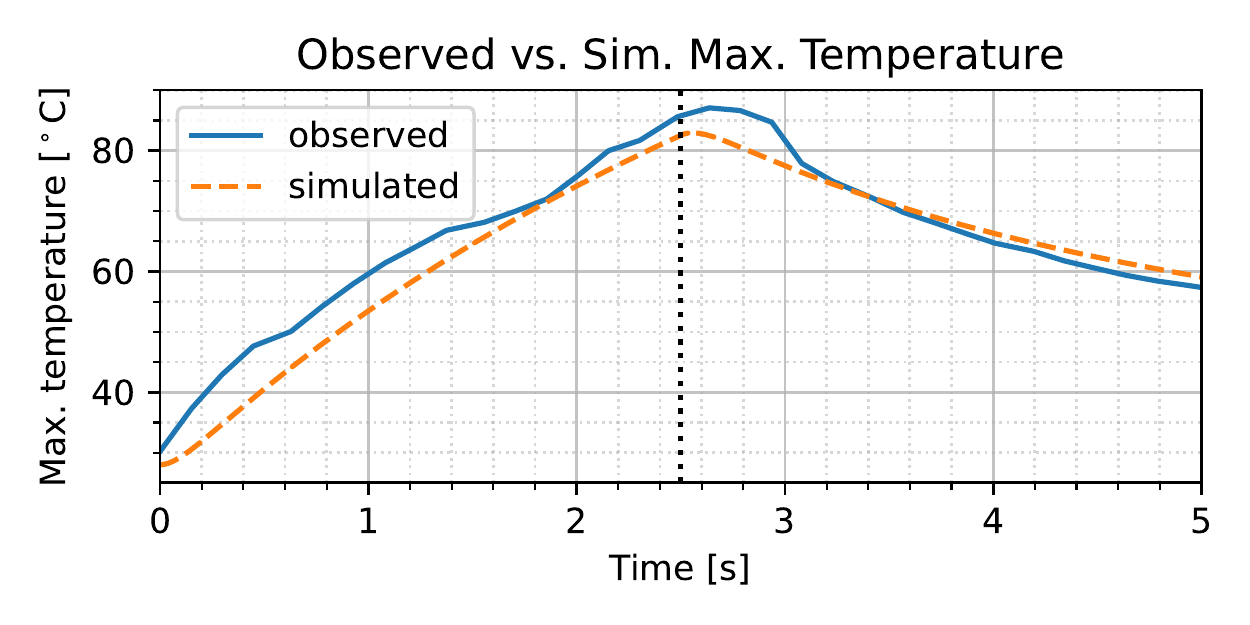}
    \caption{Maximum temperature measured over time for both the real-life observed thermal response, and the simulated response of porcine liver tissue subjected to a stationary 5 W electrosurgical heat input. The dotted vertical line indicates the point of probe disengagement. Clear congruence between the results can be observed, with small overall errors.}
    \label{fig:live vs sim max temperature}
\end{figure}

\begin{figure}[t]
\begin{minipage}[b]{0.49\linewidth}
  \centering
  \centerline{\includegraphics[width=\linewidth, trim=4.3cm 0 0.5cm 0, clip]{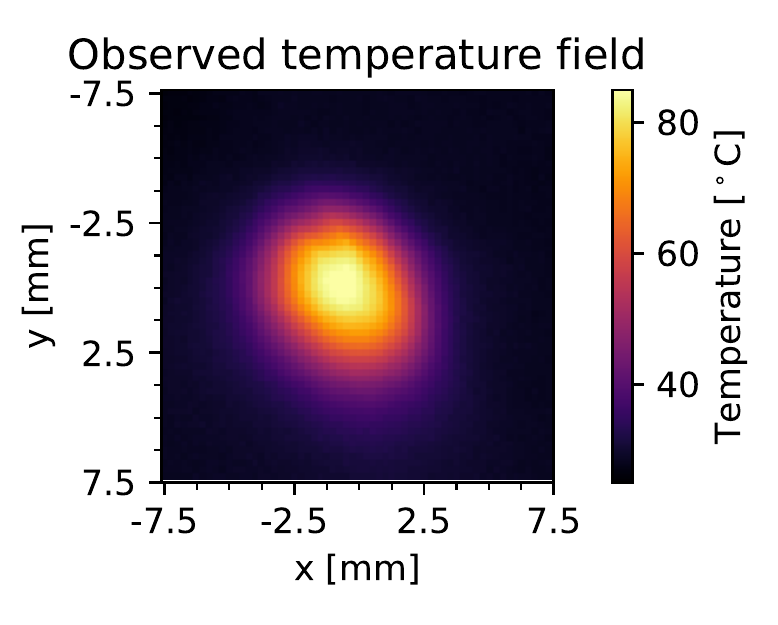}}
  \centerline{\small (a) Meas. temperature field.}
\end{minipage}
\hfill
\begin{minipage}[b]{0.49\linewidth}
  \centering
  \centerline{\includegraphics[width=\linewidth, trim=4.3cm 0 0.5cm 0, clip]{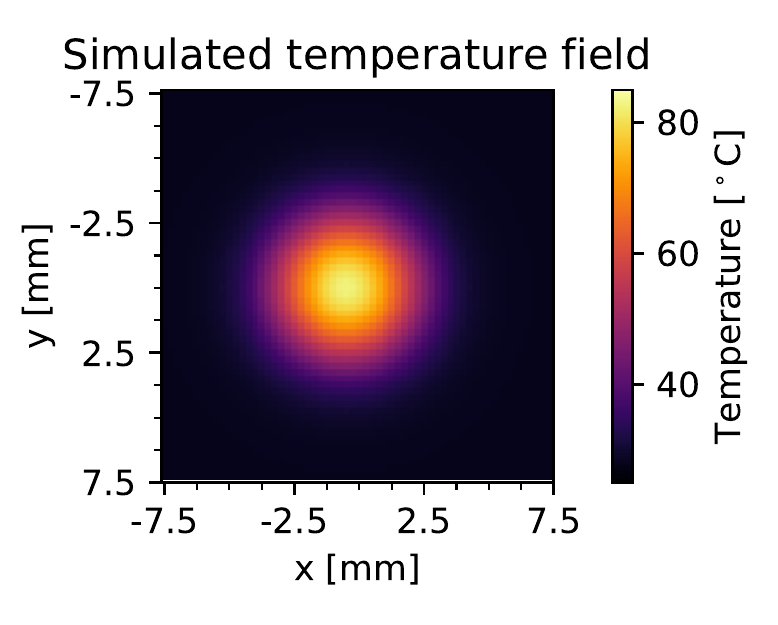}}
  \centerline{\small (b) Sim. temperature field.}
\end{minipage}
    \caption{Comparison between the temperature field at the point of probe disengagement, for both (a) the measured real-life data, as well as (b) the simulated data.}
    \label{fig:max temperature field comparison}
\end{figure}

We rerun the simulation using the parameter estimates obtained in the foregoing; we have chosen the initial temperature to be 28 $^\circ$C, which is the average temperature prior to engaging the electrosurgical unit. To illustrate the agreement between the observed real-life thermal response, and the simulated response, the maximum temperature time series plotted in Fig.~\ref{fig:live vs sim max temperature} provides a clear overview. We note that given the parameters obtained by application of our method, the simulated maximum temperature discrepancy is under 5 $^\circ$C  (6\%) with respect to real-life data across the entire simulation run. To provide further evidence of the parameter fidelity, we compare the temperature fields obtained at 2.5 seconds, when the probe is disengaged. In Fig.~\ref{fig:max temperature field comparison} the observed and simulated temperature fields are shown. Most of the error can be attributed to misalignment of the camera and the liver tissue's curvature, but despite these effects, the maximum temperature error is less than 12 $^\circ$C  (14\%). We have previously addressed mismatches between simulation and sensor measurements in detail, proving that an observer structure with output feedback can eliminate these discrepancies \cite{El-Kebir2022c}.


\section{Conclusion}\label{sec:conclusion}

We have presented a novel approach to real-time parameter estimation for a hyperbolic heat transfer model using thermographic data, as applied to \emph{in vivo} porcine liver tissue subjected to electrosurgical action. Unlike past methods, which require the solution of complex PDE systems, our approach, based on our previous work on \emph{attention-based noise robust averaging} (ANRA) \cite{El-Kebir2022c}, operates directly on thermographic data and requires only the PDE structure to be given. 
We have applied our approach to a theoretical test case involving a Maxwell--Cattaneo model with parameters based on porcine liver tissue, as well as real-life data from an \emph{in vivo} porcine liver subject to electrosurgical action. In all cases, our method produces parameter estimates close to the ground truth or parameters reported in the literature. Using the results obtained by our method, we have also presented a comparison between live and dead tissue thermal behavior.
In addition, a simulation was run using these new parameter estimates, achieving clear correspondence with the real-life porcine liver thermal response. With this validation, we have shown that \emph{our method is capable of in situ real-time minimally invasive high-fidelity thermophysical parameter estimation}.

Some of the limitations of this work, planned to be addressed in our future research, arise due to the challenging live tissue geometry and breathing motion, and electrosurgical smoke and particles causing temperature reading distortion. The future work may extend this approach to other thermodynamic models, such as the reaction-diffusion equation. In addition, we plan to develop methods to discern tissue types in real-time based on their thermal response. Related to this is planned work on comparing tissue properties in \emph{ex vivo} and \emph{in vivo} tissue, with the goal of deriving robust relationships between the two. We also plan to further investigate thermal shock waves and their significance to energy-based surgery and the tissue thermal response.

\section*{Animal Use Approval Details}

The animal use protocols (ACC NO: 19-151 and 22-097) were reviewed in accordance with the Animal Care Policies and Procedures of the University of Illinois at Chicago. Original protocol approval: 9/24/2019 (3 year approval with annual continuation required). Current approval period: 8/9/2022 to 7/19/2025.

%
%
%

\bibliographystyle{IEEEtran}
\bibliography{root.bib}

\end{document}